\documentstyle[seceq,supplement,epsf]{ptptex}

\pubinfo{No. 148, 2002}  
\notypesetlogo  

\title{
Hoop Conjecture and Black Holes on a Brane
}

\author{
Kouji {\sc Nakamura}$^{1,}$\footnote{E-mail: kouchan@th.nao.ac.jp}, 
Ken-ichi {\sc Nakao}$^{2,}$\footnote{E-mail: knakao@sci.osaka-cu.ac.jp}
and 
Takashi {\sc Mishima}$^{3,}$\footnote{E-mail: tmishima@ns.phys.ge.cst.nihon-u.ac.jp}
}

\inst{
$^1$Division of Theoretical Astrophysics, National Astronomical
Observatory, \\
Mitaka 181-8588, Japan
\\
$^2$Department of Physics, Graduate School of Science, Osaka
City University, \\
Osaka 558-8585, Japan
\\
$^3$Laboratories of Physics, College of Science and Technology,
Nihon University, \\
 Funabashi 274-0063, Japan
}

\recdate{
April 30, 2002}

\abst{
The initial data of gravity for a cylindrical matter distribution
confined to a brane are studied in the framework of the single-brane 
Randall-Sundrum scenario. 
In this scenario, the 5-dimensional nature of gravity appears in the
short-range gravitational interaction. 
We find that a sufficiently thin configuration of matter leads to
the formation of a marginal surface, even if the configuration is
infinitely long. 
This implies that the hoop conjecture proposed by Thorne does not hold
on the brane: Even if a mass $M$ does not become compacted into a region
whose circumference ${\cal C}$ in every direction satisfies 
${\cal C}> 4\pi GM$, black holes with horizons can form in the
Randall-Sundrum scenario.
}

\begin{document}

\maketitle

\section{Introduction}
\label{sec:intro}

{\it Black hole horizons form when and only when a mass $M$ gets
compacted into a region whose circumference ${\cal C}$ in every
direction is ${\cal C}< 4\pi GM$.} 
This is the statement of Thorn's hoop conjecture for the necessary and
sufficient condition for black hole formation in 4-dimensional general
relativity.\cite{Ref:HOOP}
Assuming physically reasonable conditions on the matter fields, Thorne
has proven that there is no marginal surface in a system with a
cylindrical distribution of matter fields.\cite{Ref:HOOP} 
In his proof, the marginal surface is a cylindrically symmetric
space-like 2-surface such that the expansion of the outgoing null
normal to this surface vanishes. 
This result, together with the Newtonian analogy, led to his
conjecture.


Because of the ambiguities in the above statement, there are many
proposals for a precise reformulation and attempts to prove this
conjecture.\cite{Ref:NAKED1,Ref:NAKED2,Ref:hoop-misc} 
Among them, the numerical simulations of Nakamura et
al.\cite{Ref:NAKED1} and Shapiro and Teukolsky\cite{Ref:NAKED2}
suggest that the hoop conjecture holds in 4-dimensional general
relativity. 
Therefore, we may assume that this conjecture gives one of the critera
for black hole formation and a highly elongated matter distribution
does not form a black holes.


However, strictly speaking, we do not know whether general relativity
can describe strong gravity in our universe even for classical
situations? 
There is no experimental evidence for it. 
If general relativity is inapplicable to the situation of the strong
gravity, it again becomes a non-trivial issue whether or not the hoop
conjecture gives a reasonable criterion of black hole formation.


Brane worlds are regarded as alternative theories of gravity. 
Among several models, one of the simplest is the single-brane
model proposed by Randall and Sundrum (RS II model).\cite{Ref:RS2}
One of the interesting features of this model is the fact that it
reproduces 4-dimensional Newtonian and general relativistic gravity on
the brane to more than adequate precision.  
The deviation in the gravitational force from the Newtonian one
appears at short length scales less than 
$l:=\sqrt{-6/\Lambda} = 3/(4\pi G_{5}\lambda)$, where $\Lambda<0$ is
the negative cosmological constant and $\lambda>0$ is the positive
tension of the brane. 
The length scale of $l$ must be less than the order of one millimeter
due to the experimental constraints.\cite{Ref:Newton-Correction}


It is natural to expect that 5-dimensional aspects of the
RS-scenario will appear in the short-range force of gravity (the scale
less than $l$), and hence in strong gravity with spacetime curvature
radius less than $l$.
Further, we can expect that the hoop conjecture applied to the RS
brane may not be valid on scales less than $l$ for the following
reason.
In the RS-scenario, the cancellation of the long-range forces due to
the negative cosmological constant and the brane tension results in
the reproduction of the Minkowski spacetime, and 4-dimensional gravity
is reproduced by the warp factor, due to the negative cosmological
constant in the bulk.  
Note that both effects are due to the cosmological constant on the
brane and bulk. 
We should note that short-range gravity is not particularly sensitive
to the cosmological constant. 
The 5-dimensional aspects of the RS II model will appear on
short-range scales.  
Since there are so-called ``black string solutions'' in 5-dimensional
Einstein gravity. 
These are cylindrically symmetric black holes in 5-dimensional
spacetime. 
Thus a highly elongated matter distribution may form due to the effect
of 5-dimensional gravity.


In this article, we confirm the above conjectures assuming the
existence of an infinite energy density of the brane. 
For simplicity, we concentrate on the time symmetric initial data,
which is a 4-dimensional space-like hypersurface embedded in the entire
spacetime with vanishing extrinsic curvature.\cite{Ref:SS}
Further, we consider a cylindrically symmetric matter distribution
on the brane and the formation of the cylindrical marginal surface on
the brane, since there is no marginal surface in this system within
4-dimensional general relativity. 
We show that a cylindrical marginal surface does form in the RS II
model.
Next, we discuss the hoop conjecture on the RS-brane.

\section{Time symmetric initial value constraint} 

In the RS II model, the entire spacetime is governed by
5-dimensional Einstein gravity.
The geometry of the 4-dimensional time-symmetric initial hypersurface
$(\Sigma,q_{ab})$ embedded in the 5-dimensional spacetime should
satisfy the Hamiltonian constraint 
\begin{equation}
  \label{eq:general-hamicon}
   {}^{(4)}\!R = 16\pi G_{5} T_{\perp\perp},
\end{equation}
where ${}^{(4)}\!R$ is the scalar curvature on $\Sigma$,
$T_{\perp\perp}:=T_{ab}u^{a}u^{b}$, $T_{ab}$ is the 5-dimensional
energy momentum tensor, and $u^{a}$ is the time-like unit vector
normal to $\Sigma$.
In the context of the RS II model, $T_{\perp\perp}$ is given by  
\begin{equation}
  T_{\perp\perp} = \frac{\Lambda}{8\pi G_{5}} 
  + \left(\lambda + \rho \right) \delta(\chi),
\end{equation}
where $\rho$ is the energy density of the matter field confined to the 
brane, and $\chi$ is the Gaussian normal coordinate of the brane.


The existence of the brane on the initial surface gives a
discontinuity of the extrinsic curvature defined by
$\kappa_{ab}:=-h_{a}^{\;\;c}h_{b}^{\;\;d}D_{c}n_{d}$, where $n_{a}$ is
the unit vector normal to ${\cal S}$, $h_{ab}:=q_{ab}-n_{a}n_{b}$, and
$D_{a}$ is the covariant derivative associated with the metric
$q_{ab}$. 
This discontinuity is derived from the initial value constraint 
(\ref{eq:general-hamicon}).
The scalar curvature on $\Sigma$ is given by 
\begin{equation}
  {}^{(4)}\!R = {}^{(3)}\!R - (\kappa^{a}_{\;\;a})^{2} 
  - \kappa_{d}^{\;\;c} \kappa_{c}^{\;\;d} 
  + 2 \frac{\partial}{\partial \chi} \kappa_{a}^{\;\;a}, 
\end{equation}
where $\chi$ is the Gaussian normal coordinate of ${\cal S}$. 
Then, the discontinuity of $\kappa^{a}_{\;\;b}$ with the
$Z_{2}$-symmetry at the brane is given by  
\begin{equation}
  \label{eqn:general-junction}
  \kappa^{a}_{\;\;a} = 4\pi G_{5}\left(\lambda + \rho\right).
\end{equation}


We assume that the intrinsic geometry on the initial data is described
by the following line element:
\begin{equation}
  d\ell^{2} := \phi^{2} d\tilde{\ell}^{2} 
  = \phi^{2} \left(dR^{2} + R^{2}d\varphi^{2} + dz^{2} + d\psi^{2}\right).  
  \label{eq:line-element}
\end{equation}
Here, $\psi$ is the coordinate for the extra dimension, which is
chosen so that the brane is at $\psi=l$ on this initial data and the
brane normal is given by $n_{a} = (d\chi)_{a} = \phi(d\psi)_{a}$. 
$(R,\varphi,z)$ represents the spatial cylindrical coordinate system on the
3-dimensional subspace orthogonal to $n_{a}$. 
The conformal factor $\phi$ is determined by the initial value
constraint (\ref{eq:general-hamicon}) and the boundary condition at
the brane (\ref{eqn:general-junction}). 
These are given by   
\begin{eqnarray}
  \label{eq:constraint-1}
  && \tilde{D}^{a}\tilde{D}_{a} \phi + \frac{1}{3} \phi^{3} \Lambda = 0,\\
  \label{eq:junction}
  && \left.\partial_{\psi} \phi + 
  \frac{4\pi G_{5}}{3} \left(\lambda + \rho\right)
  \phi^{2}\right|_{\psi=l} = 0,
\end{eqnarray}
where $\tilde{D}_{a}$ is the covariant derivative associated with the
flat line element $d\tilde{\ell}^{2}$. 
Using the fine tuning made by choosing 
$l:=\sqrt{-6/\Lambda}=3/(4\pi G_{5}\lambda)$ 
($\rho=0$), the time-symmetric hypersurface of the Minkowski brane in
the RS II model is given by $\phi=l/\psi$. 
Since we concentrate on the cylindrically symmetric matter
distribution on the brane, we solve Eq.~(\ref{eq:constraint-1}) under
the boundary condition (\ref{eq:junction}) with $\rho=\rho(R)$.


\section{Cylindrical marginal surfaces on the brane}

As explained in Ref.9), we numerically solved
(\ref{eq:constraint-1}). 
To solve (\ref{eq:constraint-1}), we consider the energy density  
\begin{equation}
  \rho(R)\phi^{2}|_{\psi=l}={3\sigma\over \pi R_{\rm s}^{2}}
  \left\{\left({R\over R_{\rm s}}\right)^{2}-1\right\}^{2}
  \label{eq:Non-singular-rho}
\end{equation}
on the brane. 
Further, we imposed boundary conditions as follows: (i) the
junction condition (\ref{eq:junction}) at the brane; (ii) regularity
at the axis of the cylindrical radial coordinate $R=0$ by 
$\partial_{R}\phi=0$; (iii) the condition that $\phi-l/\psi$ behaves
like the linear solution $\Omega_{L} = O(G_{5}l\rho) \ll 1$ with a
singular line source $\rho(R)=\sigma_{\rm L}\delta(R)/2\pi R$ near the
numerical boundaries $\psi=\psi_{\max}$ and $R=R_{\max}$.


The boundary condition (iii) for the conformal factor $\phi$
guarantees that our cylindrical matter distribution is an isolated
system. 
The linear solution of Eq.(\ref{eq:constraint-1}) with
(\ref{eq:junction}) is given by
\begin{eqnarray}
  &&\Omega_{\rm L} := 
  \frac{2G_{5}\sigma_{\rm L}}{3} \biggl\{ \frac{3l}{\psi^{2}}
  \ln \left({R_{\rm c}\over R}\right) + 
  \int_{0}^{\infty}dmu_{m}(l)u_{m}(\psi)K_{0}(mR)\biggr\},
\label{eq:L-solution}
\end{eqnarray}
where $R_{\rm c}$ is an integration constant, $K_{0}(x)$ is the
modified Bessel function of the 0th kind, and $u_{m}(\psi)$ is a
combination of the spherical Bessel functions of the first and second
kinds: 
\begin{eqnarray}
  \label{eq:KK-mode-function}
  u_{m}(\psi) = \sqrt{\frac{2(ml)^{4}}{\pi((ml)^{2}+1)}} 
  m\psi(n_{1}(ml)j_{2}(m\psi)-j_{1}(ml)n_{2}(m\psi)).
\end{eqnarray}
Though the mode function (\ref{eq:KK-mode-function}) is slightly
different from that for the static perturbation,\cite{Ref:RS2} the
linear solution (\ref{eq:L-solution}) on the brane is reproduced to
more than adequate precision on large length scales. 
Actually, the asymptotic form of $\Omega_{\rm L}$ on the brane is
given by
\begin{equation}
  \Omega_{\rm L} \sim \frac{2G_{5}\sigma_{L}}{l} \left(
  \ln\frac{R_{c}}{R} + \frac{l^{3}}{3R^{3}} \right).
\end{equation}
The logarithmic behavior of this asymptotic form is due to the zero
mode. 
This behavior is interpreted as the asymptotically conical structure
of the initial surface, as in 4-dimensional Einstein gravity.


\begin{figure}[b]
  \begin{center}
    \leavevmode
    \epsfxsize=0.6\textwidth
    \epsfbox{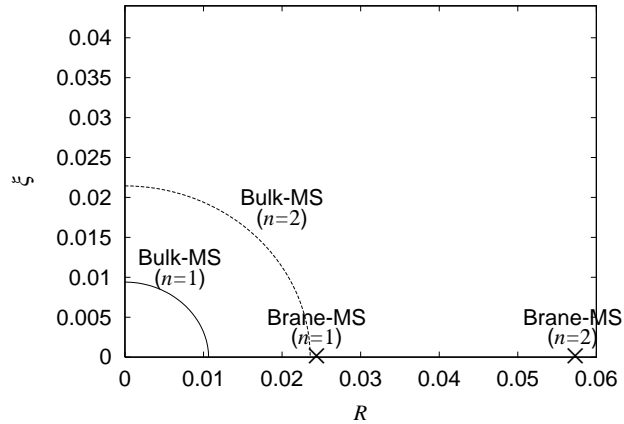}
    \caption{ The Bulk-MS and the Brane-MS are depicted. The matter is
    located within the region $0\leq R <0.01n$ at $\xi=0$, where
    $n=1,2$.} 
    \label{fig:M-surface}
  \end{center}
\end{figure}


On the initial surface with the numerical data, we found the
cylindrical marginal surfaces. 
As commented by Shiromizu and Shibata,\cite{Ref:SS} there are two
conceptually different marginal surfaces on the initial surface: One is
defined by the ``null'' rays confined to the brane ({\it Brane-MS})
and the other is defined by the null rays that propagate in the entire
spacetime, including the bulk ({\it Bulk-MS}). 
Brane-MS is the marginal surface for all the physical fields confined
to the brane, and Bulk-MS is for all the physical fields including 
gravitons that propagate in the entire spacetime. 
Brane-MS does not depend on the causal structure of the entire
spacetime, but Bulk-MS does. 
The marginal surfaces we found are shown in Fig.~\ref{fig:M-surface}.
Thus we found a cylindrical marginal surface on the brane that does
not form in 4-dimensional Einstein gravity, as Thorne proved. 
These cylindrical marginal surfaces can be regarded as counter
examples of the hoop conjecture by introducing appropriate definitions
of the mass $M$ and the circumference ${\cal C}$ in the next section.


\section{Hoop conjecture and infinite cylindrical matter distribution}

Because we found a cylindrical marginal surface in the RSII model, we
can say that the hoop conjecture is essentially violated in this
model, i.e., a black hole may form even if the matter distribution is
highly elongated. 
Further, the above examples are also regarded as counter-examples of
the hoop conjecture.  
To see this, we must introduce definitions of the mass $M$ and the
circumference ${\cal C}$ in the hoop conjecture and then we consider
the marginal surface in the situation ${\cal C} > 4\pi G M$.


As the definition of the mass, we adopt the total proper mass
\begin{equation}
  M := 2\pi \int_{0}^{R_{\rm s} } dR R \int_{-L}^{L} dz \phi^{3}(R,0) \rho(R)
  \label{eq:mass-def}
\end{equation}
within a cylinder of coordinate radius $R_{\rm s} $ and finite
coordinate length $2L$. 
As the definition of the circumference, we adopt the proper length 
\begin{equation}
  {\cal C} := 2 \int_{-L}^{L}dz\phi(R_{\rm s} ,0) 
  + 4 \int_{0}^{R_{\rm s} }dR\phi(R,0)
  \label{eq:circum-def}
\end{equation}
of this cylinder. 
We can easily check that the inequality ${\cal C}>4\pi GM$ holds for
arbitrary $L$ if and only if the following inequality is satisfied: 
\begin{equation}
  {4\pi^{2}G_{5}\over l\phi(R_{\rm s},0)}
  \int_{0}^{R_{\rm s}}dRR\phi^{3}(R,0)\rho(R)\leq1.
  \label{eq:hoop-criterion}
\end{equation}
Since the inequality (\ref{eq:hoop-criterion}) is independent of the
coordinate length $L$, we can apply this inequality to an infinite
cylinder. 
This inequality gives an upper bound on the line energy density of
the cylindrical matter field.


We also confirmed numerically that the inequality
(\ref{eq:hoop-criterion}) holds in the above examples, and these can
be regarded as counter-examples of the hoop conjecture. 
Therefore, we may say that 
{\it 
the Bulk-MS can form even for such a highly elongated matter
distribution that the inequality ${\cal C}>4\pi GM$ is satisfied.
} 
In other words, the inequality ${\cal C}> 4\pi GM$ is not a necessary
condition for the formation of black holes with horizons in the RS
scenario, although it might be a sufficient condition. 
In spite of the fact that cylindrical marginal surfaces do not exist
in 4-dimensional general relativity, our examples suggest that massive
spindle singularities in the RS scenario will be enclosed by the
marginal surface.

\section{Discussion}

In this article, we have discussed the violation of the hoop
conjecture in the RS II model. 
The existence of a cylindrical marginal surface demonstrated in this 
article is due to the 5-dimensional effect of gravity. 
Because we impose boundary conditions such that the spatial infinity
on the initial surface is flat with a finite deficit angle, we
conjecture that the existence of future null infinity similar to that
in the AdS or Minkowski spacetime and, further, that the global
hyperbolicity in the causal past of the future null infinity are
guaranteed.
If these conjectures hold, the formation of a marginal surface in the
initial data means the formation of a black hole with an event
horizon.


Though there will be static solutions with cylindrical event horizons
along the brane, we also expect that these cylindrical horizon might
be unstable, because many higher-dimensional black string solutions
are unstable.\cite{Ref:GL}
However, even though the examples above are unstable, they show that
a highly elongated marginal surface may form in the RS model as a
transient phase. 
This is one of the features of the RS model as an alternative theory
of gravity.


Though the inequality ${\cal C}> 4\pi GM$ is a sufficient condition
for the formation of black holes in RS model, as shown above, this
inequality gives an upper bound on the radius of the black string on
the brane for the following reason. 
The line energy density $\sigma$ is roughly estimated as 
$\sigma\sim M/({\cal C}/2)$, where $M$ is the mass of a portion whose
circumference is ${\cal C}$.
If the inequality ${\cal C}<4\pi GM$ holds, black holes will form in
the usual sense of 4-dimensional general relativity. 
However, when the inequality ${\cal C}>4\pi GM$ holds, the
4-dimensional effect does not produce a black hole but black holes may 
form due to the 5-dimensional effects. 
The inequality ${\cal C}>4\pi GM$ suggests 
$G_{5}\sigma < l/2\pi\sim 0.16 l$ in the RS II model. 
Because the radius of a 5-dimensional black string is also roughly
estimated by $r\sim G_{5}\sigma$, this inequality suggests that the
radius of the black string should be of order $O(0.1 l)$ or smaller.


Our examples depicted in Fig.~\ref{fig:M-surface} satisfy the upper bound
of the black string radius estimated above. 
In other words, our examples also show that on a scale a sufficiently
shorter range than $l$, black string solutions on the brane will be
approximated well by 5-dimensional black string solutions, in spite of
the existence of a timelike singular hypersurface (brane). 
Thus, in our example, the cosmological constant and the tension of the
brane do not lead to essential effects on such a scale and, the
5-dimensional features in the gravitational interaction appear as
expected. 
To clarify whether or not the above rough estimation is reasonable for
various black string solutions, it is necessary to consider black
string solutions with different thickness of matter distribution and
line energy density.


\section*{Acknowledgements}

We are grateful to D.~Ida, A.~Ishibashi, H.~Ishihara, T.~Tanaka,
G.~Uchida and H.~Kodama for their useful comments and discussions. 
This work was partially supported by a Grant-in-Aid 
for Creative Basic Research (No.~09NP0801) from the Japanese 
Ministry of Education, Science, Sports and Culture.

\end{document}